\documentclass[acmsmall, nonacm, screen]{acmart}
\acmConference[]{}
\acmPrice{}
\acmISBN{}

\renewcommand\footnotetextcopyrightpermission[1]{}
\setcopyright{none}
\AtBeginDocument{%
  }

\usepackage{bussproofs}

\newcommand{\etylizer}{\emph{etylizer}}

\newcommand{\one}{\mathbb{1}}
\newcommand{\zero}{\mathbb{0}}
\newcommand{\base}{b}
\newcommand{\const}{c}
\newcommand{\bases}{\mathbb{B}}
\newcommand{\types}{\mathcal{T}}
\newcommand{\typescirc}{\types^\bigcirc}
\newcommand{\Etypes}{\mathbb{E}\mathcal{T}}
\newcommand{\Etypescirc}{\Etypes^\bigcirc}
\newcommand{\variables}{\mathcal{V}}
\newcommand{\fields}{\mathcal{F}}

\newcommand{\map}[1]{\pmb{\texttt{\{}}#1\pmb{\texttt{\}}}}

\newcommand{\subty}{\leq}

\newcommand{\domain}{\mathcal{D}}

\newcommand{\domainbot}{\domain_\Omega}
\newcommand{\interpret}[1]{[\![#1]\!]}
\newcommand{\interpretbot}[1]{\interpret{#1}_\Omega}
\newcommand{\Einterpret}[1]{\mathbb{E}(#1)}
\newcommand{\Finterpret}[1]{\mathscr{F}(#1)}
\newcommand{\constants}{\mathcal{C}}

\newcommand{\tags}{\texttt{tags}}
\newcommand{\phimap}{\Phi_{\texttt{map}}}

\newcommand{\anydomain}{D}
\newcommand{\anydomainbot}{D_\Omega}

\newcommand{\power}{\mathcal{P}}
\newcommand{\powerfin}{\mathcal{P}_{\textnormal{fin}}}
\newcommand{\bydef}{\stackrel{\scriptsize{\textnormal{def}}}{=}}
\newcommand{\funspace}[2]{{#2}^{#1}}

\usepackage{listings}
\usepackage{verbatim}
\usepackage{xcolor}

\newcommand{\erlexp}[1]{\colorbox{black!3}{\small\texttt{#1}}}
\newcommand{\purple}[1]{\textcolor{ACMPurple}{#1}}
\newcommand{\blue}[1]{\textcolor{ACMDarkBlue}{#1}}
\newcommand{\orange}[1]{\textcolor{ACMOrange}{#1}}

\lstdefinestyle{erl}{
    breakatwhitespace=false,         
    breaklines=true,                 
    captionpos=b,                    
    keepspaces=true,                 
    numbers=left,                    
    showspaces=false,                
    showstringspaces=false,
    showtabs=false,                  
    literate =
      {-spec}{{\blue{-spec}}}5
      {-type}{{\blue{-type}}}5
      {any()}{any()}4
      {list()}{list()}5
      {map()}{map()}4
      {atom}{{\orange{atom}}}4
      {string}{{\orange{string}}}6
      {tuple}{{\orange{tuple}}}5
      {maps}{{\textcolor{black}{maps}}}3
      {:}{{\textcolor{black}{\bfseries :}}}1
      {function}{{\orange{function}}}8
      {nested}{{\orange{nested}}}5
      {A}{{\purple{A}}}1
      {F}{{\purple{F}}}1
      {G}{{\purple{G}}}1
      {T}{{\purple{T}}}1
      {T1}{{\purple{T1}}}2
      {T2}{{\purple{T2}}}2
      {T3}{{\purple{T3}}}2
      {K}{{\purple{K}}}1
      {K1}{{\purple{K1}}}2
      {K2}{{\purple{K2}}}2
      {K3}{{\purple{K3}}}2
      {V}{{\purple{V}}}1
      {V1}{{\purple{V1}}}2
      {V2}{{\purple{V2}}}2
      {V3}{{\purple{V3}}}2
      {Y}{{\purple{Y}}}1
      {X}{{\purple{X}}}1
      {X1}{{\purple{X1}}}2
      {X2}{{\purple{X2}}}2
      {As}{{\purple{As}}}2
      {As1}{{\purple{As1}}}3
      {As2}{{\purple{As2}}}3
      {Ks}{{\purple{Ks}}}2
      {\_Ks}{{\purple{\_Ks}}}3
      {Grp}{{\purple{Grp}}}3
      {Map}{{\purple{Map}}}3
      {Map1}{{\purple{Map1}}}4
      {Map2}{{\purple{Map2}}}4
      {Map3}{{\purple{Map3}}}4
      {Tree}{{\purple{Tree}}}4
      {List}{{\purple{List}}}4
      {Tree1}{{\purple{Tree1}}}5
      {Tree2}{{\purple{Tree2}}}5
      {Fun}{{\purple{Fun}}}3
      {Acc}{{\purple{Acc}}}3
      {Init}{{\purple{Init}}}4
      {Default}{{\purple{Default}}}7
      {Combiner}{{\purple{Combiner}}}7
 }
\lstdefinelanguage{erlang}{%
   morekeywords={
       after,case,catch,div,end,exit,export,if,import,module,of,or,%
       receive,rem,throw,when,fun%
   },
   numbers=left,
   frame=lines,
   otherkeywords={},%
   xleftmargin=0.0cm,
   numbersep=5pt,
   sensitive=true,%
   columns=,
   commentstyle=\color{ACMLightBlue},
   keywordstyle=\color{ACMDarkBlue},
   stringstyle=\color{ACMPurple},
   basicstyle=\ttfamily\small,
   numberstyle=\tiny\color{ACMDarkBlue!75}\sffamily\raisebox{0.6pt},
   morecomment=[n]{\{-}{-\}},%
   morestring=[b]"%
  }[keywords,comments,strings]%

\newcommand{\DefTypes}{
    \begin{define}[Types]\label{Def:types}
	Let $\types$ be the set of terms $t$ generated coinductively by the following grammar such that (1) $t$ has a finite number of different subterms (regularity) and (2) every infinite branch in $t$ contains an infinite number of occurrences of the map type constructor (contractivity).
	\begin{align*}
	    \begin{split}
		& \text{\bfseries Type} \hspace{2ex}
		t \hspace{1ex} \textnormal{::=} 
		    \hspace{1ex} \alpha \hspace{1ex} 
		    | \hspace{1ex} \base \hspace{1ex}
		    | \hspace{1ex} \map{f,...,f} \hspace{1ex}
		    | \hspace{1ex} t \lor t \hspace{1ex}
		    | \hspace{1ex} \neg t \hspace{1ex}
		    | \hspace{1ex} \zero
		\\
		& \text{\bfseries Field} \hspace{1ex}
		f \hspace{1ex} 
		\textnormal{::=} 
		\hspace{1ex} t = t \hspace{1ex} 
		  | \hspace{1ex} t \Rightarrow t
	    \end{split}
	\end{align*}
	Here, $\base$ ranges over a set of base types $\bases$, and $\alpha$ over some countable set of type variables $\variables$.
	As the set of types we define $\typescirc$ that also contains markings $t^=,t^\Rightarrow$ for any $t\in\types$:
	\begin{displaymath}
	    \typescirc\bydef\{t,t^=,t^\Rightarrow \ | \ t\in\types\}. 
	\end{displaymath}
    \end{define}
}

\newcommand{\DefInterpretation}{
    \begin{define}[Set-theoretic Interpretation] 
	A \emph{set-theoretic interpretation} of types $\typescirc$ is given by a set $\anydomain$ and a function 
	$\interpret{\cdot}:\typescirc\to\power(\anydomainbot)$ such that
	for any $t,t_1,t_2\in\types$
    \begin{displaymath}
	\interpret{t_1\lor t_2} = \interpret{t_1}\cup\interpret{t_2}
	\hspace{4ex}
	\interpret{\neg t} = \anydomain\setminus\interpret{t}
	\hspace{4ex}
	\interpret{\zero} = \emptyset
    \end{displaymath}
    and 
    \begin{displaymath}
	\interpret{t^=} = \interpret{t}
	\hspace{4ex}
	\interpret{t^\Rightarrow} = \interpretbot{t}.
    \end{displaymath}
    \end{define}
}

\newcommand{\DefFunctionSpace}{
    \begin{define}[Function Space]
	Let $X,Y$ be sets. A function $f:X\to Y$ is a subset of $X\times Y$ such that it is
	\begin{displaymath}
	    (\forall x\in X.(!\exists y\in Y. (x,y)\in f)).
	\end{displaymath}
	Any subset of $\funspace{X}{Y}$ is called a \emph{function space} where
	\begin{displaymath}
	    \funspace{X}{Y} \bydef
	    \{f\subseteq X\times Y \ | \ f:X\to Y \}.
	\end{displaymath}
    \end{define}
}
\newcommand{\DefFunctionSpaceMerge}{
    \begin{define}[Merge]
	Let $(\funspace{X_i}{Y_i})_{i\in I}$ be a finite family of function spaces. Then their \emph{merge} is defined as
	\begin{displaymath}
	    \sum_{i\in I}(\funspace{X_i}{Y_i}) \bydef 
	    \bigl\{\bigcup_{i\in I}f_i \ | \ \forall i\in I. f_i\in
	    (\funspace{X_i}{Y_i})\bigl\}
	\end{displaymath}
	if all $X_i$ are pairwise disjoint, otherwise it is defined as
	\begin{displaymath}
	    \sum_{i\in I}(\funspace{X_i}{Y_i}) \bydef 
	    \sum_{I'\subseteq I}
	    \funspace
	    {\Bigl(\bigcap_{i\in I'}X_i\setminus\bigcup_{i\in I\setminus I'}X_i\Bigl)}
	    {\Bigl(\bigcup_{i\in I'}Y_i\Bigl)}.
	\end{displaymath}
	Furthermore, let $(\funspace{X_j}{Y_j})_{j\in J}$ be another finite family of funtion spaces. Then the merge of merged function spaces is defined as
	\begin{displaymath}
	    \sum_{i\in I}(\funspace{X_i}{Y_i}) +
	    \sum_{j\in J}(\funspace{X_j}{Y_j}) \bydef
	    \sum_{k\in (I\cup J)}(\funspace{X_k}{Y_k}).
	\end{displaymath}
    \end{define}
}

\newcommand{\DefExtensionalInterpretation}{
    \begin{define}[Extensional Interpretation]
	Let $\interpret{\cdot}:\typescirc\to\power(\anydomainbot)$ be a set-theoretic interpretation. Then its associated \emph{extensional interpretation} is the unique set-theoretic interpretation 
	$\Einterpret{\cdot}:\Etypescirc\to\power(\mathbb{E}\anydomainbot)$ 
	such that it is
	\begin{align*}
	    \Einterpret{b} &= \bases(b)\subseteq\constants \\
	    \Einterpret{\map{f_1,...,f_r}} &=
	    \Finterpret{f_1}+\cdots+\Finterpret{f_r}\subseteq
	    \funspace{\anydomain}{\anydomainbot}
	\end{align*}
	where 
	\begin{align*}
	    \Etypescirc &= \{t,t^=,t^\Rightarrow \ | \ t\in\Etypes\} \\
	    \Etypes &=\types\setminus\{\map{f_1,...,f_r}\in\types \ | \ 
	    \Finterpret{f_1}+\cdots+\Finterpret{f_r}
	    \not\subseteq\funspace{\anydomain}{\anydomainbot}\} \\
	    \mathbb{E}\anydomainbot &= \constants+\anydomainbot+
	    \funspace{\anydomain}{\anydomainbot}.
	\end{align*}
	The function 
	$\Finterpret{\cdot}:\fields\to\power
	(\power(\anydomain\times\anydomainbot))$ 
	is called the \emph{field interpretation} and defined as
	\begin{displaymath}
	    \Finterpret{t_1 = t_2} \bydef
	    \funspace{\interpret{t_1}}{\interpret{t^=_2}}
	    \hspace{4ex}
	    \Finterpret{t_1 \Rightarrow t_2} \bydef
	    \funspace{\interpret{t_1}}{\interpret{t^\Rightarrow_2}}.
	\end{displaymath}
    \end{define}
}
\newcommand{\DefModel}{
    \begin{define}[Model]
	A set-theoretic interpretation 
	$\interpret{\cdot}:\typescirc\to\power(\anydomainbot)$ is a 
	\emph{model} if it induces the same containment relation as its associated extensional interpretation:
	\begin{displaymath}
	    \forall t_1,t_2\in(\Etypescirc).
	    \interpret{t_1}\subseteq\interpret{t_2}\iff
	    \Einterpret{t_1}\subseteq\Einterpret{t_2}.
	\end{displaymath}
    \end{define}
}
\newcommand{\DefDomain}{
    \begin{define}[Interpretation Domain]\label{Def:domain}
	The \emph{interpretation domain} $\domain$ of types is the set of finite terms $d$ generated inductively by the following grammar
	\begin{align*}
		d \hspace{1ex} \textnormal{::=} \hspace{1ex} 
		\const^V \hspace{1ex}
		| \hspace{1ex} \{(d,\delta),...,(d,\delta)\}^V \hspace{1ex}
		\hspace{4ex}
		\delta \hspace{1ex} \textnormal{::=} \hspace{1ex}
		d \hspace{1ex} 
		| \hspace{1ex} \Omega
	\end{align*}
	where $\const$ ranges over constants $\constants$ and $V$ ranges over sets of variables contained in $\variables$.
    \end{define}
}

\newcommand{\DefSubtypingRelation}{
    \begin{define}[Subtyping]
	Let $\interpret{\cdot}:\typescirc\to\power(\domainbot)$ be the interpretation from Corollary \ref{Cor:model}. 
	Then subtyping is defined over 
	$\Etypescirc\times\Etypescirc$ as the set containment
	\begin{displaymath}
	    t_1\subty t_2 \stackrel{\textnormal{def}}{\iff}
	    \interpret{t_1}\subseteq\interpret{t_2}.
	\end{displaymath}
    \end{define}
}

\newcommand{\DefSubstitutions}{
    \begin{define}
	A \emph{type substitution} $\sigma:\variables\to\types$ is a mapping from type variables to types, such that $\sigma$ is the identity everywhere except on a finite set of variables. This finite set is called the domain of the substitution and denoted as
	$\texttt{dom}(\sigma)=\{\alpha \ | \ \sigma(\alpha)\neq\alpha \}$.
	We write $t\sigma$ for the application of $\sigma$ to the type $t\in\types$.
    \end{define}
}

\newcommand{\CorollaryModel}{
    \begin{corollary}[Well-founded Model]\label{Cor:model}
	The following definition of 
	$\interpret{\cdot}:\types^\bigcirc\to\power(\domainbot)$ is a well-founded model:
	\begin{displaymath}
	    \forall t\in\types.\interpret{t}\bydef
	    \{d\in\domain \ | \ (d:t)\}.
	\end{displaymath}
	Here, the binary predicate $(d:t)$ over $\domain\times\types$ is defined by induction on the pair $(d,t)$ ordered lexicographically as
	\begin{align*}
	    (d:\alpha) &= \alpha\in\tags(d) \\
	    (\const^V:\base) &= \const\in\bases(\base) \\
	    \bigl(\{(d,\delta)\}^V:\map{t_1 = t_2}\bigl) &= 
	    (d:t_1)\land(\delta:t_2)\land(\delta\neq\Omega) \\
	    \bigl(\{(d,\delta)\}^V:\map{t_1 \Rightarrow t_2}\bigl) &= 
	    (d:t_1)\land((\delta:t_2)\lor(\delta=\Omega)) \\
	    \bigl(\{(d_i,\delta_i) \ | \ i\in I \}^V:\map{f_1,...,f_r}\bigl) &=
	    \forall i\in I.\exists j\in\{1,...,r\}.
	    \bigl(\{(d_i,\delta_i)\}^V:\map{f_j}\bigl) \\
	    (d: t_1\lor t_2) &= (d:t_1)\lor(d:t_2) \\
	    (d:\neg t) &= \neg(d:t) \\
	    (d:t) &= \texttt{false} \textnormal{ in any other case}.
	\end{align*}
    \end{corollary}
}

\newcommand{\PropositionSimpleContainment}{
    \begin{proposition}\label{Prop:simple-containment}
	For finite $I$, following holds for merged function spaces:
	\begin{displaymath}
	    \sum_{i\in I}(\funspace{X_i}{Y_i})
	    \subseteq 
	    \power\Bigl(\bigcup_{i\in I} X_i\times Y_i\Bigl).
	\end{displaymath}
    \end{proposition}
}

\newcommand{\PropositionSimpleMergeContainment}{
    \begin{proposition}\label{Prop:simple-merge-containment}
	Containment problem of merged function spaces can be reduced to containment problem of power sets and vice versa:
	\begin{displaymath}
	    \sum_{i\in I}(\funspace{X_i}{Y_i})
	    \subseteq
	    \sum_{j\in J}(\funspace{X_j}{Y_j})
	    \iff
	    \power\Bigl(\bigcup_{i\in I} X_i \times Y_i \Bigl)
	    \subseteq
	    \power\Bigl(\bigcup_{j\in J} X_j \times Y_j \Bigl).
	\end{displaymath}
	where $I,J$ finite and it is $\bigcup_{i\in I}X_i = \bigcup_{j\in J}X_j$.
    \end{proposition}
}

\newcommand{\TheoremMergeIntersection}{
    \begin{theorem}[Merge Intersection]\label{Th:intersection}
	Let $\mathcal{I} = \{I_1,...,I_p\}$ be a family of index sets. Then the following holds:
	\begin{displaymath}
	    \bigcap_{I\in\mathcal{I}}\sum_{i\in I}(\funspace{X_i}{Y_i}) =
	    \sum_{(i_1,...,i_p)\in I_1\times\cdots\times I_p}
	    \funspace
	    {\bigl(X_{i_1}\cap\cdots\cap X_{i_p} \bigl)}
	    {\bigl(Y_{i_1}\cap\cdots\cap Y_{i_p} \bigl)}.
	\end{displaymath}
    \end{theorem}
}
\newcommand{\TheoremMergeContainment}{
    \begin{theorem}[Merge Containment]\label{Th:merge-containment}
	Let $(I_p)_{p\in P}$ and $(J_n)_{n\in N}$ be finite families of finite index sets. Then the following holds:
	\begin{displaymath}
	    \bigcap_{p\in P}\sum_{i\in I_p}(\funspace{X_i}{Y_i})
	    \subseteq
	    \bigcup_{n\in N}\sum_{j\in J_n}(\funspace{X_j}{Y_j})
	    \iff
	    \bigcap_{p\in P}\power\Bigl(\bigcup_{i\in I_p} X_i \times Y_i \Bigl)
	    \subseteq
	    \bigcup_{n\in N}\power\Bigl(\bigcup_{j\in J_n} X_j \times Y_j 
	    \Bigl).
	\end{displaymath}
	where it is $\bigcup_{i\in I_p}X_i = \bigcup_{j\in J_n}X_j$ for all $p\in P$ and $n\in N$.
    \end{theorem}
}

\newcommand{\TheoremMapContainment}{
    \begin{theorem}[Map Containment]\label{Th:map-containment}
	Let $P,N\subseteq\Etypes$ be finite sets of map types. Then it is
	\begin{displaymath}
	    \bigcap_{t\in P}\Einterpret{t}\subseteq
	    \bigcup_{t\in N}\Einterpret{t}
	    \iff
	    \bigcap_{\map{f_1,...,f_r}\in P}
	    \power\Bigl(\bigcup^r_{i=1}X_i\times Y_i\Bigl)\subseteq
	    \bigcup_{\map{f_1,...,f_r}\in N}
	    \power\Bigl(\bigcup^r_{i=1}X_i\times Y_i\Bigl)
	\end{displaymath}
	where $X_i,Y_i$ come from $\funspace{X_i}{Y_i}=\Finterpret{f_i}$.
    \end{theorem}
}
\newcommand{\TheoremMapIntersection}{
    \begin{theorem}[Map Intersection]\label{Th:map-intersection}
	Let $\emptyset\neq P\subseteq\Etypes$ be a finite set of map types. Then exists some $\map{f_1,...,f_r}\in\Etypes$ such that
	\begin{displaymath}
	    \bigcap_{t\in P}\Einterpret{t} =
	    \Einterpret{\map{f_1,...,f_r}}.
	\end{displaymath}
    \end{theorem}
}

\newcommand{\CorollarySubtypingAlgorithm}{
    \begin{corollary}\label{Cor:subtyping}
	Let $\tau=\map{t_1\circ_1 s_1,...,t_p\circ_p s_p}\in\Etypes$ and $N\subseteq\Etypes$ a finite set of map types. Then the following definition of a subtyping decomposition algorithm (for $\tau\subty\bigvee_{t\in N}t$) is sound and complete
	\begin{align*}
	    \phimap(\tau, N)=
	    \exists & \map{u_1\bullet_1 v_1,...,u_n\bullet_n v_n}\in N. \\
	    &\hspace{3ex}\forall i\in\{1,...,p\}. 
	    \Bigl(
		\forall J\subseteq\{1,...,n\}.
		\bigl(t_i\subty\bigvee_{j\in J}u_j\bigl)
		\hspace{1ex}\textnormal{or}\hspace{1ex}
		\bigl(s_i^{\circ_i}\subty
		\bigvee_{j\in\{1,...,n\}\setminus J}v^{\bullet_j}_j\bigl)
	    \Bigl)
	\end{align*}
	where $\circ_i,\bullet_j\in\{=,\Rightarrow\}$.
    \end{corollary}
}

\newcommand{\PropSubstitution}{
    \begin{proposition} If $t_1\subty t_2$ then $t_1\sigma\subty t_2\sigma$ for any type substituiton $\sigma$.
    \end{proposition}
}

\newcommand{\ProofSimpleMergeContainment}{
    \begin{proof}
    It is
    \begin{align*}
	\begin{split}
	    \sum_{i\in I}(\funspace{X_i}{Y_i})
	    &\subseteq
	    \sum_{j\in J}(\funspace{X_j}{Y_j}) 
	    \\
	    &\iff
	\sum_{k\in(I\cup J)}(\funspace{X_k}{Y_k}) = 
	\sum_{j\in J}(\funspace{X_j}{Y_j})
	    \\
	    &\iff
	    \power\Bigl(\bigcup_{k\in(I\cup J)} X_k \times Y_k \Bigl) =
	    \power\Bigl(\bigcup_{j\in J} X_j \times Y_j \Bigl)
	    \\
	    &\iff
	    \power\Bigl(\bigcup_{i\in I} X_i \times Y_i \Bigl) \subseteq
	    \power\Bigl(\bigcup_{j\in J} X_j \times Y_j \Bigl).
	\end{split}
    \end{align*}
    \end{proof}
}

\newcommand{\ProofIntersection}{
    \begin{proof}\hfill

    \noindent
    $(\subseteq):$
    Let $f\in\bigcap_I\sum_{i}(\funspace{X_i}{Y_i})$.
    For each $I\in\mathcal{I}$ it is
    \begin{displaymath}
	f = \bigcup_{\emptyset\neq I'\subseteq I}g_{I'} =: G_I
	\hspace{8ex}
	g_{I'}\in
	\funspace
	{\bigl(\bigcap_{i\in I'}X_i\setminus\bigcup_{i\in I\setminus I'}X_i\bigl)}
	{\bigl(\bigcup_{i\in I'}Y_i\bigl)}
	=: A_{I'}.
    \end{displaymath}
    By rewriting $f = \bigcap_I G_I$ we get
    \begin{displaymath}
	f = \bigcup_{\emptyset\neq(I'_1\times\cdots\times I'_p)\subseteq(I_1\times\cdots\times I_p)}
	    g_{I'_1}\cap\cdots\cap g_{I'_p}.
    \end{displaymath}
    Moreover, each union member is an element of $A_{I'_1}\cap\cdots\cap A_{I'_p} = \funspace{X}{Y}$ with
    \begin{align*}
	\begin{split}
	    X &= \Bigl(\bigcap_{(i_1,...,i_p)\in I'_1\times\cdots\times I'_p}
	    X_{i_1}\cap\cdots\cap X_{i_p}\Bigl)
	    \setminus
	    \bigcup_{
		(i_1,..,i_p)\in(I_1\times\cdots\times I_p)\setminus
		(I'_1\times\cdots\times I'_p)
	    }X_{i_1}\cap\cdots\cap X_{i_p}
	    \\
	    Y &= \bigcup_{(i_1,...,i_p)\in I'_1\times\cdots\times I'_p}Y_{i_1}\cap\cdots\cap Y_{i_p}.
	\end{split}
    \end{align*}
    Lastly, definition of merge implies
    \begin{displaymath}
	f\in\sum_{\emptyset\neq(I'_1\times\cdots\times I'_p)\subseteq(I_1\times\cdots\times I_p)}\bigl(A_{I'_1}\cap\cdots\cap A_{I'_p}\bigl)
	= \sum_{(i_1,...,i_p)\in I_1\times\cdots\times I_p}
	    \funspace
	    {\bigl(X_{i_1}\cap\cdots\cap X_{i_p} \bigl)}
	    {\bigl(Y_{i_1}\cap\cdots\cap Y_{i_p} \bigl)}.
    \end{displaymath}
    \par\noindent
    $(\supseteq):$ Apply above proof backwards.
    \end{proof}
}

\newcommand{\ProofMergeContainment}{
   \begin{proof}
   Let us define the shorthand 
   \begin{displaymath}
       \mathcal{M}_K := \sum_{k\in K}(\funspace{X_k}{Y_k}) 
   \end{displaymath}
   for each index set $K\in (I_p)_p\cup(J_n)_n$.
   \par\noindent
   $(\Rightarrow):$ First, show that there exists some $n\in N$ satisfying
   $\mathcal{M} := \bigcap_p \mathcal{M}_{I_p}
   \subseteq \mathcal{M}_{J_n}$.
   Assume there is no such $n$. Then, there are functions $f_1,...,f_{|N|}\in\mathcal{M}$ such that
   \begin{displaymath}
       f_i\in(\mathcal{M}\cap\mathcal{M}_{J_{n_i}})\setminus
       \bigcup_{n\in N\setminus\{n_i\}}\mathcal{M}_{J_n}
   \end{displaymath}
   for $i=1,...,|N|$ and $n_i\in N$. Now consider functions
   \begin{displaymath}
       g_i\in\mathcal{M}_{J_{n_i}}\setminus
       \bigcup_{n\in N\setminus\{n_i\}}\mathcal{M}_{J_n}
   \end{displaymath}
   for which it is $f_i\cap g_i\neq\emptyset$. At this point, for each $i=1,...,|N|$ there is some element $(x_i,y_i)\in f_i\cap g_i$. Observe that it is $(x_i,y_i)\in h\in\mathcal{M}$ for some function $h$. The fact that $h\not\in\mathcal{M}_{J_n}$ for all $n\in N$ contradicts the initial assumption that 
   $\mathcal{M}\subseteq\bigcup_{n\in N}\mathcal{M}_{J_n}$.
   \par
   Thus exists some $n\in N$ satisfying $\mathcal{M}\subseteq\mathcal{M}_{J_n}$.
   After rewriting $\mathcal{M}$ to some $\mathcal{M'}$ by using theorem \ref{Th:intersection} and then applying proposition \ref{Prop:simple-merge-containment} on $\mathcal{M'}\subseteq\mathcal{M}_{J_n}$ results in
   \begin{displaymath}
       \bigcap_{p\in P}\power\Bigl(\bigcup_{i\in I_p}X_i\times Y_i\Bigl)
	\subseteq
       \bigcup_{n\in N}\power\Bigl(\bigcup_{j\in J_n}X_j\times Y_j\Bigl).
   \end{displaymath}
   \par\noindent
   $(\Leftarrow):$ Without limitations $P = \{1,...,p\}$. We have
   \begin{displaymath}
      \bigcap_{p\in P}\power\Bigl(\bigcup_{i\in I_p}X_i\times Y_i\Bigl)
       = \power\Bigl(\bigcup_{(i_1,...,i_p)\in I_1\times\cdots\times I_p}(X_{i_1}\cap\cdots\cap X_{i_p})\times(Y_{i_1}\cap\cdots\cap Y_{i_p})\Bigl) =: A.
   \end{displaymath}
   Moreover, exists some $n\in N$ satisfying
   \begin{displaymath}
       A\subseteq\power\Bigl(\bigcup_{j\in J_n}X_j\times Y_j\Bigl).
   \end{displaymath}
   Applying proposition \ref{Prop:simple-merge-containment} results in
   \begin{displaymath}
       \sum_{(i_1,...,i_p)\in I_1\times\cdots\times I_p}
       \funspace
       {\bigl(X_{i_1}\cap\cdots\cap X_{i_p}\bigl)}
       {\bigl(Y_{i_1}\cap\cdots\cap Y_{i_p}\bigl)}
       \subseteq
       \sum_{j\in J_n}\funspace{X_j}{Y_j}.
   \end{displaymath}
   With theorem \ref{Th:intersection} applied to the left-hand side, we get:
   \begin{displaymath}
       \bigcap_{p\in P}\sum_{i\in I_p}(\funspace{X_i}{Y_i})
       \subseteq
       \bigcup_{n\in N}\sum_{j\in J_n}(\funspace{X_j}{Y_j}).
   \end{displaymath} 
   \end{proof}
}

\newcommand{\ProofSubtypingAlgorithm}{
    It is
    \begin{align*}
	\tau\subty\bigvee_{t\in N}t&\iff
	\interpret{\tau}\subty\bigcup_{t\in N}\interpret{t} \\
	&\iff\powerfin\Bigl(
	    \bigcup^p_{i=1}\interpret{t_i}\times\interpret{s^{\circ_i}_i}
	    \Bigl)\subseteq
	    \bigcup_{\map{u_1\bullet_1 v_1,...,u_n,\bullet_n v_n}\in N}
	    \powerfin\Bigl(
		\bigcup^n_{i=1}\interpret{u_i}\times\interpret{v^{\bullet_i}_i}
		\Bigl) \\
	&\iff\exists\map{u_1\bullet_1 v_1,...,u_n\bullet_n v_n}\in N.
	    \Bigl(\bigcup^p_{i=1}\interpret{t_i}\times\interpret{s^{\circ_i}_i}
	    \subseteq
	    \bigcup^n_{i=1}\interpret{u_i}\times\interpret{v^{\bullet_i}_i}
	    \Bigl)\\
	&\iff\exists\map{u_1\bullet_1 v_1,...,u_n\bullet_n v_n}\in N.
	    \forall i\in\{1,...,p\}.
	    \Bigl(
		\interpret{t_i}\times\interpret{s^{\circ_i}_i}\subseteq
	    \bigcup^n_{i=1}\interpret{u_i}\times\interpret{v^{\bullet_i}_i}	
		\Bigl) \\
	&\iff\exists\map{u_1\bullet_1 v_1,...,u_n\bullet_n v_n}\in N. \\
	&\hspace{10ex}\forall i\in\{1,...,p\}.
	\Bigl(
	    \forall J\subseteq\{1,...,n\}.\bigl(
		\interpret{t_i}\subseteq\bigcup_{j\in J}\interpret{u_j}\bigl)
		\hspace{1ex}\textnormal{or}\hspace{1ex}
		\bigl(
		\interpret{s^{\circ_i}_i}\subseteq
		\bigcup_{j\in\{1,...,n\}\setminus J}\interpret{v^{\bullet_j}_j}
		\bigl)
	\Bigl) \\
	&\iff\phimap(\tau, N).
    \end{align*}
}

\newcommand{\ResultMapsMoreGeneral}{
    \begin{corollary}
	Let $\interpret{\cdot}:\typescirc\to\power(\domainbot)$ be the interpretation from Corollary \ref{Cor:model}.
	Let $(\map{t_i = s_i,\ \neg t_i\Rightarrow\one})_{i\in I}$ be a finite family of map types. Then exists some 
	$\map{f_1,...,f_r}\in\Etypes$ such that
	    \begin{displaymath}
		\bigcap_{i\in I}
		\interpret{\map{t_i = s_i,\ \neg t_i\Rightarrow\one}}=
		\interpret{\map{f_1,...,f_r}}.
	    \end{displaymath}
    \end{corollary}
}

\newcommand{\ProofMapsMoreGeneral}{
    \begin{proof} Consequence of theorem \ref{Th:map-intersection}. 
    \end{proof}
}

\AtEndPreamble{
    \theoremstyle{acmplain}
    \newtheorem{define}[theorem]{Definition}
}

\DeclareMathAlphabet{\mathcal}{OMS}{cmsy}{m}{n}

\begin{document}
\title{Semantic Subtyping for Maps in Erlang}

\author{Erdem Yildirim}
\affiliation{%
  \institution{University of Kaiserslautern-Landau}
  \country{Germany}}
\email{e_yildirim19@cs.uni-kl.de}

\author{Albert Schimpf}
\orcid{0009-0009-4172-5730}
\affiliation{%
  \institution{University of Kaiserslautern-Landau}
  \country{Germany}}
\email{schimpf@cs.uni-kl.de}

\author{Stefan Wehr}
\orcid{0000-0001-5242-767X}
\affiliation{%
  \institution{Offenburg Univ. of Applied Sciences}
  \country{Germany}}
\email{stefan.wehr@hs-offenburg.de}

\author{Annette Bieniusa}
\orcid{0000-0002-1654-6118}
\affiliation{%
  \institution{University of Kaiserslautern-Landau}
  \country{Germany}}
\email{bieniusa@cs.uni-kl.de}

\renewcommand{\shortauthors}{Yildirim et al.}

\begin{abstract}
    In this paper we will construct a set-theoretic model of types featuring type variables, base types, set-theoretic types and map types. Syntax of map types spans all the map types available in Erlang. The model of types is used to define a semantic subtyping relation based on set containment. The novelty of this work is the definition of subtyping over parameteric map types.
\end{abstract}

\ccsdesc[300]{Theory of computation~Type structures}
\ccsdesc[300]{Software and its engineering~Functional languages}

\keywords{set-theoretic types, maps, erlang}


\maketitle

\section{Introduction}
In this work we will introduce a semantic subtyping relation for all of the map types in Erlang. Since those map types are allowed to be parametric in key domain as well as in value domain, the novelty of this work is the definition of subtyping over parametric map types. 

We obtain the subtyping relation by interpreting (parametric) map types as subsets of an abstract domain representing all values of the language. Then, subtyping is defined as set containment over those type interpretations.




We see this work as an extension of \cite{C23} where a semantic subtyping relation is given for non-parametric map types, that is map types without type variables. We are also using the same theory as \cite{C23}, called \emph{semantic subtyping}, to obtain our subtyping relation.

There are two usages of maps as summarized in \cite{C23}. Further, the author shows that $(1)$ is a special case of $(2)$, so that both can be typed in a unified way.
\begin{enumerate}
    \item \emph{Maps as records:} In these maps, all keys are known statically before program execution. This is ensured by restricting possible keys to some predefined domain which can be different than the values of the language.
    \item \emph{Maps as dictionaries:} In these maps there are no restrictions on the key domain. Thus keys can be values or expressions that compute values. This implies that not all keys are known statically, instead they are computed during program execution. 
\end{enumerate}

This work is motivated by the flexibility of map expressions and map types present in Erlang, where programming with maps is augmented with parametric map types and pattern matching. The authors have realized that the typing of Erlang maps cannot be done in the system of \cite{C23}.

The authors would like to express that this work is the first part of a bigger work. The second part will be about typing of map expressions in Erlang using the typechecker \etylizer\ of \cite{SWB23}.

\subsection{Polymorphic Function Specifications}
In Erlang, top-level functions can be annotated by types. These types are called specifications. As an example, consider the following specifications: 

\begin{lstlisting}[firstnumber=1,linerange={7-8}]
-spec merge(#{K1 => V1}, #{K2 => V2}) -> #{K1 => V1, K2 => V2}.
-spec intersect(#{K1 => term()}, #{K2 => V2}) -> #{ety:and(K1,K2) => V2}.
\end{lstlisting}
\noindent
Here, the type $\erlexp{ety:and/2}$ is provided by \etylizer\ and represents a set-theoretic intersection of types. Both $\erlexp{merge/2}$ and $\erlexp{intersect/2}$ are part of the standard library. Observe that this kind of typing is more precise than the following possible typing:

\begin{lstlisting}[firstnumber=3,linerange={2-3}]
-spec merge(map(), map()) -> map().
-spec intersect(map(), map()) -> map().
\end{lstlisting}
\noindent
This precision is achieved by conserving key and value types of the inputs to the output.

As another example, consider the following specifications:

\begin{lstlisting}[firstnumber=5,linerange={50-51}]
-spec remove(K, #{K1 => V1}) -> #{K1 => V1}.
-spec remove(K, #{K1 => V1}) -> #{K => none(), K1 => V1}.
\end{lstlisting}
\noindent
As can be seen, typing of functions operating on maps can be ambiguous. This ambiguity is a source of confusion when formalizing map types. In our formalization, for example, both of the output types are given the same interpretation.

Moreover, not every typing is precise enough. Consider the following specification:

\begin{lstlisting}[firstnumber=7,linerange={9-9}]
-spec merge(#{K1 => V1}, #{K2 => V2}) -> #{K1|K2 => V1|V2}.
\end{lstlisting}
\noindent
Here, the output type uses type unions in contrast to the output in line $1$.
We argue that both typings in lines $1$ and $7$ should be correct. In our formalization, we have the output type of line $1$ as a proper subtype of the output type in line $7$. Thus, line $1$ is a more precise typing than line $7$.

\subsection{Polymorphic Function Definitions}
Let us consider two different definitions of the function $\erlexp{with/2}$ from maps standard library. This function takes a list $\erlexp{\purple{Ks}}$ and a map $\erlexp{\purple{Map}}$, then returns a new map containing those key-value pairs of $\erlexp{\purple{Map}}$ whose keys are in the given list. The first definition uses $\erlexp{intersect/2}$ whose type was given earlier in line $2$:

\begin{lstlisting}[firstnumber=8,linerange={42-46}]
-spec with([K], #{K1 => V1}) -> #{ety:and(K,K1) => V1}.
with(Ks, Map) -> 
   maps:intersect(
      lists:foldr(fun maps:merge/2, #{}, [#{K => true} || K <- Ks]),
      Map).
\end{lstlisting}
\noindent
Since the specification of $\erlexp{intersect/2}$ contains an intersection type, above typing of $\erlexp{with/2}$ may seem as a simple deduction. However, what we ideally want to have is to deduce the intersection type from more complex situations. Hence, consider the second definition that uses pattern matching on maps:

\begin{lstlisting}[firstnumber=13,linerange={33-40}]
-spec with([K], #{K1 => V1}) -> #{ety:and(K,K1) => V1}.
with(Ks, Map) -> 
   lists:foldr(fun(K, Acc) -> 
      case Map of 
        #{K := V} -> maps:merge(#{K => V}, Acc);
	#{} -> Acc
      end
   end, #{}, Ks).
\end{lstlisting}
\noindent

In order to typecheck these definitions w.r.t. its specification we want to use the \etylizer\ type system. As we have said before, typing of maps will be the subject of the second part of our work.

\subsection{Overview}
In section \ref{Sec:types} we present our integration of parametric map types into semantic subtyping theory. To this end, the syntax of types is introduced and a set-theoretic model of types $\interpret{\cdot}$ is constructed. This model interprets map types as subsets of $\powerfin(\domain\times\domainbot)$ where 
$\domain$ is the interpretation domain and $\domainbot=\domain\cup\{\Omega\}$. It is proven that such an interpretation of maps results in the same containment relation as if maps denoted subsets of the function space $\funspace{\domain}{\domainbot}$. This space consists of those functions $f:\domain\to\domainbot$ which are defined on all the points in $\domain$. 
The "$\Omega$" stands for those mappings $d\mapsto\Omega$ in $f$ which mimics the absence of the element $d$.

Following this, the subtyping relation is defined as set containment over the model $\interpret{\cdot}$ and accordingly a subtyping algorithm is given for parametric map types.

Lastly, type substitutions are defined and it is proven that subtyping is preserved by substitutions.

\section{Types}\label{Sec:types}
We will define the syntax of types and construct a set-theoretic model for them that is well-founded. Then we will define a subtyping algorithm for parametric map types that is sound and complete w.r.t. the model. We will also prove that subtyping is preserved by type substitutions. Lastly, we will show that our formalization is a generalization of the so called \emph{arrow types} in semantic subtyping theory.

\subsection{Syntax of Types}
\DefTypes
We assume that there exists the set $\constants$ of language constants, and the function $\bases(\cdot):\bases\to\power(\constants)$ maps base types to sets of constants. We further assume that $\bases$ includes \emph{singleton types} $\base_\const$ for each constant $\const\in\constants$ such that $\bases(\base_\const)=\{c\}$. Singleton types are necessary to model record types. For example, the map type
\begin{displaymath}
    \map{\base_\texttt{name}=\textnormal{String},\ 
    \base_\texttt{age}=\textnormal{Int}}
\end{displaymath}
models records having attributes \texttt{name} and \texttt{age} mapped to strings and integers.

%

\subsection{Set-theoretic Model of Types}
\subsubsection{Rationale}We assume that map types $\map{f_1,...,f_r}$ do not impose any field ordering. To emphasize this fact, we will define a commutative operation "$+$" over sets such that the extensional interpretation satisfies 
\begin{displaymath}
    \Einterpret{\map{f_1,...,f_r}}=\Finterpret{f_1}+\cdots+\Finterpret{f_r}
\end{displaymath}
for a field interpretation $\Finterpret{\cdot}$. Fields will be interpreted as function spaces:
\begin{align*}
    \Finterpret{t=s} = \funspace{\interpret{t}}{\interpret{s}}
    \hspace{4ex}
    \Finterpret{t\Rightarrow s} = \funspace{\interpret{t}}{\interpretbot{s}}.
\end{align*}
Moreover, we will use the interpretation domain $\domain=\constants+\powerfin(\domain\times\domainbot)$ of \cite[Section 5.2]{GGL15} for polymorphic set-theoretic types $\types$ where we define our model $\interpret{\cdot}:\types^\bigcirc\to\power(\domainbot)$ by giving fixed interpretations to type variables\footnote{An equivalent model is also given by \cite{CX11} where type variables lack a fixed interpretation.}. The set $\types^\bigcirc$ additionally contains marked types $t^=,t^\Rightarrow$ for each type $t\in\types$. This marking is relevant for the subtyping algorithm, otherwise it will stay as a negligible detail in formalization.

The subtyping relation we use will be the one induced by $\interpret{\cdot}:$
\begin{displaymath}
    t\subty s \iff \interpret{t}\subseteq\interpret{s}.
\end{displaymath}

The $\Einterpret{\cdot}$ will also be called the \emph{extensional interpretation} associated to $\interpret{\cdot}$.
In order to use $\interpret{\cdot}$ as a model, we will prove an equivalence between two containment relations 
\begin{displaymath}
    (\subseteq_1)\subseteq(\funspace{\domain}{\domainbot})^2
    \hspace{4ex}
    (\subseteq_2)\subseteq\powerfin(\domain\times\domainbot)^2.
\end{displaymath}
This proof is needed for two reasons:
\begin{itemize}
    \item to reduce the values of
	$\Einterpret{\map{f_1,...,f_r}}\subseteq\funspace{\domain}{\domainbot}$
	onto values in
	$\interpret{\map{f_1,...,f_r}}\subseteq\powerfin(\domain\times\domainbot)$;
    \item to obtain a \emph{sound} and \emph{complete} subtyping algorithm for map types that depends solely on tuple decomposition.
\end{itemize}

\subsubsection{Interpretation}
\DefInterpretation

\subsubsection{Function Spaces}
\DefFunctionSpace
\DefFunctionSpaceMerge
\PropositionSimpleContainment
\begin{proof} Choose arbitrary $f_i\subseteq X_i\times Y_i$ for each $i\in I$. Observe that $\bigcup_i f_i \in\power(\bigcup_i X_i\times Y_i)$, which immediatley implies the proposition.
\end{proof}

\subsubsection{Extensional Interpretation}
\DefExtensionalInterpretation
\TheoremMapIntersection
\begin{proof} See appendix \ref{Subsec:proof-intersection}.
\end{proof}
\TheoremMapContainment
\begin{proof} See appendix \ref{Subsec:proof-map-containment}.
\end{proof}

\subsubsection{Model Construction}
\DefModel
\DefDomain
\CorollaryModel
\begin{proof} Proof is under construction.
\end{proof}


\subsection{Subtyping}
\DefSubtypingRelation
\CorollarySubtypingAlgorithm
\begin{proof} See appendix \ref{Subsec:proof-algorithm}.
\end{proof}

\subsection{Substitutions}
The upcoming description of the \etylizer\ type system uses type substitutions to instantiate type variables. We briefly introduce the notion of substitution.
\DefSubstitutions
Application of a type substitution $t\sigma$ satisfies the following equalities
\begin{align*}
    \alpha\sigma &= \sigma(\alpha) & \zero\sigma = \zero \\
    \base\sigma  &= \base	   & (t_1\lor t_2)\sigma = t_1\sigma\lor t_2\sigma \\
    \map{t_1\circ_1 s_1,...,t_r\circ_r s_r}\sigma 
    &= \map{t_1\sigma\circ_1 s_1\sigma,...,t_r\sigma\circ_r s_r\sigma} &
    (\neg t)\sigma = \neg(t\sigma)
\end{align*}
where $\circ_i\in\{=,\Rightarrow\}$. It follows that type substitutions preserve subtyping.
\PropSubstitution
\begin{proof} Proof is under construction.
\end{proof}

\subsection{An Important Result}\label{Subsec:important-result}
Our formalization yields the result that parametric map types arise as a generalization of the so called arrow types formalized elsewhere in the semantic subtyping theory.

\ResultMapsMoreGeneral
\ProofMapsMoreGeneral

The reader can check that in our model
\begin{align*}
    \interpret{\map{t=s,\ \neg t\Rightarrow\one}}&=
    \powerfin\Bigl(\overline{\interpret{t}\times\overline{\interpret{s}}^{\domainbot}}^{\domain\times\domainbot}\Bigl) & (\star) \\
    \interpret{\map{t\Rightarrow s,\ \neg t\Rightarrow\one}}&=
    \powerfin\Bigl(\overline{\interpret{t}\times\overline{\interpret{s}}^{\domain}}^{\domain\times\domainbot}\Bigl)
\end{align*}
where $(\star)$ is the standard interpretation of the arrow type $t\to s$ in semantic subtyping.

\section{Related Work}\label{Sec:related-work}
\paragraph{Typing maps}
The work most resembling to ours is  \cite{C23}, in which the author presents a unified typing for records, maps and structs in a type theory including union, intersection and negation types coupled with a subtyping relation. The author uses the theory of semantic subtyping to define the subtyping relation. Elementary results on semantic subtyping can be found in \cite{FCB08} and \cite{CX11}.

Semantic interpretation given to map types in \cite{C23} have two drawbacks. The first one being that map types lack parameterized key domains. That is, type variables cannot occur as key types. We belive that the given quasi $\mathbb{K}$-step function semantics cannot be extended reasonably to encapsulate key domain parametricity. The second drawback is that the whole key domain is partitioned into disjoint (predefined) types and this partitioning is fixed for each map type.
These two drawbacks are the reasons why we do not use quasi $\mathbb{K}$-step function semantics of maps.

Aside from mentioned drawbacks, the subtyping algorithm of \cite{C23} is more efficient than ours. This is because our algorithm checks containment between 
$(i)$ a power set of union of tuples and $(ii)$ a union of power sets of union of tuples. This problem is inherently more complex than checking containment between $(i)$ a tuple and $(ii)$ a union of tuples, which the algorithm in \cite{C23} checks.

Furthermore, the author conjectures in \cite{C23} that the interpretation of map types, without restricting their key domain to some fixed partition, will be the same as arrow types. Section \ref{Subsec:important-result} of our formalization validates this conjecture.

\paragraph{Row polymorphism}
Another parallel work to ours is \cite{CP24}. There the authors restrict themselvses to the records usage of maps and extends the work of \cite{C23} by integrating row polymorphism into semantic subtyping theory. Their semantic interpretation of records is based on (partial) quasi-constant functions which are a special case of the more general quasi $\mathbb{K}$-step functions. As before, the immediate drawback is that type variables cannot occur as key types. Another drawback is that record keys, called labels, are not first-class values of the language. This means, labels cannot be computed by expressions.

Aside from its drawbacks, our experience tells that row polymorphic records are more precise in typing those map operations that are defined on individual keys.
Such operations include, for example, selection/deletion/update of statically known keys. In contrast to this, parametric map types are most useful for typing operations that are defined on the whole map structure instead of individual keys. Operations of this kind use recursion to either construct a map bottom-up or destructively traverse a map top-down. As example, have a look at the $\erlexp{with/2}$ function we have given in the introduction.


\section{Conclusion and Future Work}\label{Sec:conclusion}
We have defined a subtyping relation for a type algebra featuring semantic subtyping on type variables, basic types, parametric map types and set-theoretic types. Our map types cover all the Erlang maps. The next step of this work is already in construction. We are in the process of formally integrating parametric map types into the \etylizer\ type system. To this end, we are $(i)$ defining dynamic semantics for map patterns and expressions, $(ii)$ introducing typing rules for map patterns and expressions, and $(iii)$ extending the typing algorithm of \etylizer\ which is to be done by adding constraint generation and solving rules. 

We shall express that solely an integration to \etylizer\ is not enough to typecheck maps in Erlang. One must tackle the problem of typing maps standard library, where not every function has a definition written down in Erlang. For example, $\erlexp{merge/2}$ and $\erlexp{put/2}$ are defined as \emph{native interface functions} (nif) calling C code under the hood. Also some functions as $\erlexp{fold/3}$ and $\erlexp{to\_list/1}$ contain some Erlang code but they ultimately make calls to $\erlexp{erts\_internal}$ module containing nifs. To typecheck maps standard library we plan to $(i)$ implement each of the functions in Erlang, and then $(ii)$ write down a more precise specification than the current one (if possible).

As another line of future work, we would like to assess the efficiency of our map subtyping algorithm.


\bibliographystyle{ACM-Reference-Format}
\bibliography{main}


\begin{thebibliography}{6}


\ifx \showCODEN    \undefined \def \showCODEN     #1{\unskip}     \fi
\ifx \showISBNx    \undefined \def \showISBNx     #1{\unskip}     \fi
\ifx \showISBNxiii \undefined \def \showISBNxiii  #1{\unskip}     \fi
\ifx \showISSN     \undefined \def \showISSN      #1{\unskip}     \fi
\ifx \showLCCN     \undefined \def \showLCCN      #1{\unskip}     \fi
\ifx \shownote     \undefined \def \shownote      #1{#1}          \fi
\ifx \showarticletitle \undefined \def \showarticletitle #1{#1}   \fi
\ifx \showURL      \undefined \def \showURL       {\relax}        \fi
\providecommand\bibfield[2]{#2}
\providecommand\bibinfo[2]{#2}
\providecommand\natexlab[1]{#1}
\providecommand\showeprint[2][]{arXiv:#2}

\bibitem[Castagna(2023)]%
        {C23}
\bibfield{author}{\bibinfo{person}{Giuseppe Castagna}.}
  \bibinfo{year}{2023}\natexlab{}.
\newblock \showarticletitle{Typing Records, Maps, and Structs}.
\newblock \bibinfo{journal}{\emph{Proc. ACM Program. Lang.}}
  \bibinfo{volume}{7}, \bibinfo{number}{ICFP}, Article \bibinfo{articleno}{196}
  (\bibinfo{date}{Aug.} \bibinfo{year}{2023}), \bibinfo{numpages}{44}~pages.
\newblock
\href{https://doi.org/10.1145/3607838}{doi:\nolinkurl{10.1145/3607838}}


\bibitem[Castagna and Peyrot(2025)]%
        {CP24}
\bibfield{author}{\bibinfo{person}{Giuseppe Castagna} {and}
  \bibinfo{person}{Lo\"{\i}c Peyrot}.} \bibinfo{year}{2025}\natexlab{}.
\newblock \showarticletitle{Polymorphic Records for Dynamic Languages}.
\newblock \bibinfo{journal}{\emph{Proc. ACM Program. Lang.}}
  \bibinfo{volume}{9}, \bibinfo{number}{OOPSLA1}, Article
  \bibinfo{articleno}{132} (\bibinfo{date}{April} \bibinfo{year}{2025}),
  \bibinfo{numpages}{28}~pages.
\newblock
\href{https://doi.org/10.1145/3720497}{doi:\nolinkurl{10.1145/3720497}}


\bibitem[Castagna and Xu(2011)]%
        {CX11}
\bibfield{author}{\bibinfo{person}{Giuseppe Castagna} {and}
  \bibinfo{person}{Zhiwu Xu}.} \bibinfo{year}{2011}\natexlab{}.
\newblock \showarticletitle{Set-theoretic foundation of parametric polymorphism
  and subtyping}.
\newblock \bibinfo{journal}{\emph{SIGPLAN Not.}} \bibinfo{volume}{46},
  \bibinfo{number}{9} (\bibinfo{date}{Sept.} \bibinfo{year}{2011}),
  \bibinfo{pages}{94–106}.
\newblock
\showISSN{0362-1340}
\href{https://doi.org/10.1145/2034574.2034788}{doi:\nolinkurl{10.1145/2034574.2034788}}


\bibitem[Frisch et~al\mbox{.}(2008)]%
        {FCB08}
\bibfield{author}{\bibinfo{person}{Alain Frisch}, \bibinfo{person}{Giuseppe
  Castagna}, {and} \bibinfo{person}{V\'{e}ronique Benzaken}.}
  \bibinfo{year}{2008}\natexlab{}.
\newblock \showarticletitle{Semantic subtyping: Dealing set-theoretically with
  function, union, intersection, and negation types}.
\newblock \bibinfo{journal}{\emph{J. ACM}} \bibinfo{volume}{55},
  \bibinfo{number}{4}, Article \bibinfo{articleno}{19} (\bibinfo{date}{Sept.}
  \bibinfo{year}{2008}), \bibinfo{numpages}{64}~pages.
\newblock
\showISSN{0004-5411}
\href{https://doi.org/10.1145/1391289.1391293}{doi:\nolinkurl{10.1145/1391289.1391293}}


\bibitem[Gesbert et~al\mbox{.}(2015)]%
        {GGL15}
\bibfield{author}{\bibinfo{person}{Nils Gesbert}, \bibinfo{person}{Pierre
  Genev\`{e}s}, {and} \bibinfo{person}{Nabil Laya\"{\i}da}.}
  \bibinfo{year}{2015}\natexlab{}.
\newblock \showarticletitle{A Logical Approach to Deciding Semantic Subtyping}.
\newblock \bibinfo{journal}{\emph{ACM Trans. Program. Lang. Syst.}}
  \bibinfo{volume}{38}, \bibinfo{number}{1}, Article \bibinfo{articleno}{3}
  (\bibinfo{date}{Oct.} \bibinfo{year}{2015}), \bibinfo{numpages}{31}~pages.
\newblock
\showISSN{0164-0925}
\href{https://doi.org/10.1145/2812805}{doi:\nolinkurl{10.1145/2812805}}


\bibitem[Schimpf et~al\mbox{.}(2023)]%
        {SWB23}
\bibfield{author}{\bibinfo{person}{Albert Schimpf}, \bibinfo{person}{Stefan
  Wehr}, {and} \bibinfo{person}{Annette Bieniusa}.}
  \bibinfo{year}{2023}\natexlab{}.
\newblock \showarticletitle{Set-theoretic Types for Erlang}. In
  \bibinfo{booktitle}{\emph{Proceedings of the 34th Symposium on Implementation
  and Application of Functional Languages}} (Copenhagen, Denmark)
  \emph{(\bibinfo{series}{IFL '22})}. \bibinfo{publisher}{Association for
  Computing Machinery}, \bibinfo{address}{New York, NY, USA}, Article
  \bibinfo{articleno}{4}, \bibinfo{numpages}{14}~pages.
\newblock
\showISBNx{9781450398312}
\href{https://doi.org/10.1145/3587216.3587220}{doi:\nolinkurl{10.1145/3587216.3587220}}


\end{thebibliography}

\appendix

\section{PROOFS}
\subsection{Proof of Map Intersection}\label{Subsec:proof-intersection}
We prove the following more general theorem.
\TheoremMergeIntersection
\ProofIntersection

\subsection{Proof of Map Containment}\label{Subsec:proof-map-containment}
We prove the following more general theorem.
\TheoremMergeContainment
\ProofMergeContainment

\subsection{Proof of Simple Containment}\label{Subsec:proof-simple-merge-containment}
\PropositionSimpleMergeContainment
\ProofSimpleMergeContainment

\subsection{Proof of Corollary \ref{Cor:subtyping}}\label{Subsec:proof-algorithm}
\ProofSubtypingAlgorithm

\end{document}